\documentclass[aps,showpacs,twocolumn,showkeys,amsmath,amssymb,floatfix]{revtex4}
\usepackage{graphicx}% Include figure files
\usepackage{dcolumn}% Align table columns on decimal point
\usepackage{bm}% bold math
\usepackage{mathrsfs}
\usepackage{epsfig}

\newcommand{\N}{N_c}

\begin{document}

\title{Are There Tetraquarks at Large $N_c$ in QCD(F)?}

\author{Thomas D. Cohen}
\email{cohen@physics.umd.edu}
\affiliation{Department of Physics, University of Maryland, College Park,
Maryland 20742-4111, USA}

\author{Richard F. Lebed}
\email{richard.lebed@asu.edu}
\affiliation{Department of Physics, Arizona State University, Tempe,
Arizona 85287-1504, USA}

\date{March, 2014}

\begin{abstract}
  Weinberg recently pointed out a flaw in the standard argument that
  large $N_c$ QCD with color-fundamental quarks [QCD(F)] cannot yield
  narrow tetraquark states.  In particular, he observed that the
  argument does not rule out narrow tetraquarks associated with the
  leading-order connected diagrams; such tetraquarks would have a
  width scaling as $N_c^{-1}$.  It is shown here, however, that while
  the standard analysis of tetraquarks does not rule them out, a more
  thorough analysis rules out quantum-number exotic tetraquarks
  associated with the leading-order connected diagrams.  This analysis
  is based entirely on conventional assumptions used in large $N_c$
  physics applied to the analytic properties of meson-meson
  scattering.  Our result implies that one of three possibilities must
  be true: i) quantum-number exotic tetraquarks do not exist at large
  $N_c$; ii) quantum-number exotic tetraquarks exist, but are
  associated with subleading connected diagrams and have anomalously
  small widths that scale as $N_c^{-2}$ or smaller; or iii) the
  conventional assumptions used in large $N_c$ analysis are
  inadequate.
% This is in contrast to the case of large $N_c$ QCD with
% quarks in color two-index antisymmetric quarks [QCD(AS)].  QCD(AS)
%  at large $N_c$ can be shown using the conventional assumptions to
%  have quantum-number exotic tetraquarks arising from the lowest order
%  connected graphs.
\end{abstract}

\pacs{11.15.Pg, 12.39.Mk, 14.40.Rt}
% 11.15.Pg Expansions for large numbers of components (e.g.,
%	1/Nc expansions) 
% 12.39.Mk Glueball and nonstandard multi-quark/gluon states
% 14.40.Rt Exotic mesons

\keywords{tetraquark, large $\N$ QCD}
\maketitle

\section{Introduction} \label{sec:Intro}

Long-accepted conventional wisdom~\cite{Witten:1979kh,Coleman}
precludes the large $N_c$ limit of QCD~\cite{'tHooft:1973jz} from
supporting narrow tetraquark states ({\it i.e.}, whose widths are
suppressed by powers of $1/N_c$).  The original argument was
formulated by Witten~\cite{Witten:1979kh} and described in Coleman's
classic lectures on large $N_c$ physics~\cite{Coleman}.  Recently
Weinberg~\cite{Weinberg:2013cfa} has questioned this lore by pointing
out a serious loophole in the reasoning.

The argument espoused by Witten~\cite{Witten:1979kh} and
Coleman~\cite{Coleman} states that a tetraquark source of the form
$J=\overline{q}q\overline{q}q$ can always be expressed (owing to Fierz
reordering) as a sum of products of two color-singlet $\overline{q}q$
operators, each of which sources single-meson states at leading order
in $1/N_c$.  Therefore, the two-point function of $J$, which is
$O(N_c^2)$ due to the two closed, disconnected color loops that follow
the quark and antiquark lines, is dominated by the creation and
annihilation of two-meson states.  Since each meson terminus supplies
a decay constant $f_M = O(N_c^{1/2})$, the scaling of the full
two-point function is saturated by the $O(N_c^0)$ free propagation of
two noninteracting mesons and no tetraquark states.

However, as pointed out in Ref.~\cite{Weinberg:2013cfa}, this argument
only applies to the leading-order ($N_c^2$) disconnected diagram; the
leading {\em connected} diagram has only a single quark color loop
[$O(N_c^1)$] and therefore excludes the leading order of free
two-meson propagation.  For the connected contribution of the $J$
two-point function, the propagation of single free tetraquark states
with decay constants $f_T = O(N_c^{1/2})$ can contribute at leading
order.  Moreover, analysis~\cite{Weinberg:2013cfa} of the connected
three-point function of a $J$ source and two bilinear $\overline{q}q$
sources $B$, which is also $O(N_c^1)$, gives a tetraquark-meson-meson
coupling of $O(N_c^{-1/2})$, producing a tetraquark decay width of
$O(N_c^{-1})$: If tetraquarks exist as $N_c \to \infty$, they are
narrow.  For certain special flavor quantum numbers, the tetraquark
states must be even narrower~\cite{Knecht:2013yqa}.

That new states should first appear in the $N_c$-subleading part of a
correlation function is not a novel idea; in fact, precisely the same
argument as above may applied to the $BBBB$ four-point correlation
function, whose leading-order ($N_c^2$) disconnected piece is again
saturated by noninteracting two-meson states, but whose leading
connected [$O(N_c^1)$] piece contains scattering through single-meson
states with $O(N_c^{-1/2})$ trilinear and $O(1/N_c)$ quartic
couplings.  This mechanism allows, for example, the well-known
resonant scattering $\pi \pi \to \rho \to \pi \pi$ to occur at large
$N_c$.  The possibility considered by Weinberg is that tetraquarks
behave similarly: If a tetraquark exists and couples to two ordinary
mesons with $O(N_c^{-1/2})$ strength, then one finds no contradiction
with standard $N_c$ counting rules for meson-meson scattering, nor
with a leading-order ($N_c^2$) disconnected contribution.  However,
obtaining $f_T \sim N_c^{1/2}$ may require a subtle limiting
mechanism~\cite{Lebed:2013aka}.

We note that certain other exotic hadrons, (those with neither
exclusively $\overline{q}q$ meson nor $qqq$ baryon quantum numbers) do
occur in the conventional large $N_c$ limit.  Glueballs have
$N_c$-suppressed couplings to mesons~\cite{Witten:1979kh}, leading
both to a parametric suppression of glueball-meson mixing amplitudes
[$O(N_c^{-1/2})$] and glueball decay widths to two-meson states
[$O(1/N_c^2)$].  Meanwhile, hybrid mesons (those requiring at least a
single ``valence gluon'' in addition to $\overline{q}q$ in order to
obtain the quantum numbers of the state, such as $J^{PC} = 1^{-+}$)
behave just like ordinary mesons at large $N_c$~\cite{Cohen:1998jb}.
%One of the outstanding mysteries of strong interaction physics is
%that, after decades of experiments, no universally accepted and
%unambiguous signal for any of these states formed of light quarks and
%gluons has yet been identified.  
The fact that hadrons with exotic quark structure often have the same
quantum numbers as nonexotic hadrons (so-called ``cryptoexotic''
states) and therefore can mix with them is a major contributing factor
to the lack of clear evidence for exotics.  The phenomenon of
cryptoexotic-conventional meson mixing is just as apparent for
large-$N_c$ tetraquarks with nonexotic quantum
numbers~\cite{Knecht:2013yqa}.  Note that, even though cleanly
identifying the role of cryptoexotics in the spectrum has intrinsic
difficulties, the literature has a long tradition of explaining
various scalar resonances that do not fit into simple quark models as
being cryptoexotics~\cite{Jaffe:1976ig,Close:2002zu,Braaten:2003he,
  Close:2003sg,Maiani:2004uc,Bignamini:2009sk,Ali:2011ug,Achasov:2012kk,
  Friedmann:2009mz}.

In this paper, we focus on tetraquark with exotic quantum numbers, and
in particular those with quantum numbers requiring at least two quarks
and two antiquarks in all Fock components.  Examples include isospin-2
states, doubly-strange states, and strange states with isospin $\frac
3 2$.  We do this to keep the issue as theoretically crisp as
possible.  The question of whether a component of a quantum state is
or is not cryptoexotic has theoretical ambiguities, and could depend
upon the precise definition of precisely what constitutes a
cryptoexotic state; such a definition may depend upon a choice of
basis or of Lorentz frame.  In contrast, quantum-number exotic
configurations are unambiguously exotic.

It is significant that the logic of Ref.~\cite{Weinberg:2013cfa} does
not rule out, but also does not guarantee, the existence of
tetraquarks.  In particular, the analysis shows that previous
arguments ruling out tetraquarks at large $N_c$ are wrong, but {\em
  not} that tetraquarks must, in fact, exist.  The analysis of
Ref.~\cite{Weinberg:2013cfa} uses the best-known $1/N_c$ expansion, in
which each quark carries a single color index in the fundamental (F)
representation of SU($N_c$), a theory we label QCD(F)\@.  However, it
is worth recalling that the extrapolation from $N_c=3$ to large $N_c$
is not unique.  For the physical case of $N_c = 3$, the color
two-index antisymmetric (AS) representation of SU($N_c$) is isomorphic
to the (anti-)fundamental via the identification $q^{ij} =
\epsilon^{ijk} q_k$.  For larger $N_c$, the two representations
differ~\cite{Corrigan:1979xf} and therefore give rise to distinct
large $N_c$ limits and distinct $1/N_c$ expansions for the two
theories.  The new expansion based on quarks in the two-index
antisymmetric representation is denoted as QCD(AS)\@.  While the study
of QCD(AS) at large $N_c$ was largely motivated by the appearance of
elegant field theoretic dualities~\cite{Armoni:2003gp,
  Armoni:2003fb,Armoni:2004uu,Armoni:2005wt}, one may instead develop
the formalism for physical
baryons~\cite{Bolognesi:2006ws,Cherman:2006iy,Cohen:2009wm} and
investigate whether QCD(F) or QCD(AS) is phenomenologically superior
for a given set of
observables~\cite{Cherman:2009fh,Lebed:2010um,Cherman:2012eg}.  In the
case of mesons, it was shown very recently~\cite{Cohen:2014via} that
the presence two distinct color indices on each QCD(AS) quark field
allows one to define a single-color-trace source $J$ with tetraquark
quantum numbers.  The implication of this finding is that tetraquarks
contribute to the $J$ two-point correlator in leading order in the
$1/N_c$ expansion but two-meson states cannot.  Therefore, tetraquarks
necessarily appear in the spectrum of QCD(AS) as narrow hadrons.

The demonstration that tetraquarks exist in QCD(AS) raises the
question of whether it is possible to go beyond Weinberg's
argument~\cite{Weinberg:2013cfa} and prove the existence of
tetraquarks in QCD(F)\@.  We have not been able to do so.  Moreover,
we suspect that the reason is likely to be that that narrow tetraquark
resonances do not exist in the large $N_c$ limit of QCD(F)\@.  This
suspicion is based on the demonstration reported here that the
scenario of Ref.~\cite{Weinberg:2013cfa}, in which tetraquarks
associated with the leading-order connected diagrams exist and have
widths of $O(1/N_c)$ (the most natural scenario if narrow tetraquarks
exist), cannot be correct for quantum-number exotic channels provided
the standard assumptions used in large $N_c$ analysis hold.  This
demonstration is based on a careful analysis of the analytic structure
of the meson-meson scattering amplitude at leading nontrivial order in
the $1/N_c$ expansion.  Ultimately, the result follows because all
connected diagrams associated with four-point functions for quark
bilinear sources at leading order in $N_c$ possess no $s$-channel cuts
associated with a tetraquark.  The result shown here implies that one
of three possibilities must be true: i) quantum-number exotic
tetraquarks do not exist at large $N_c$ for QCD(F); ii) tetraquarks
exist, but are associated with subleading connected diagrams and have
anomalously small widths that scale as $N_c^{-2}$ or smaller; or iii)
the conventional assumptions used in large $N_c$ analysis are
inadequate.

\begin{figure}
\begin{center}
\includegraphics[width=2.0in]{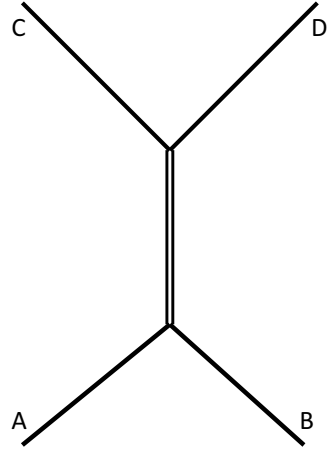}
\caption{Diagram for the $s$-channel 
tetraquark contribution to meson-meson scattering.}\label{Fig:schannel}
\end{center}
\end{figure}

The analysis is based on an indirect proof.  We begin by supposing
that the scenario proposed by Weinberg~\cite{Weinberg:2013cfa} is
correct: A narrow quantum-number exotic tetraquark resonance exists at
large $N_c$ with a coupling to mesons of $O(N_c^{-1/2})$.  Ultimately,
we will show that this assumption leads to a contradiction.  The
process at the hadronic level is indicated in Fig.~\ref{Fig:schannel}.

Before proceeding, it is useful to remind the reader that the
scattering amplitude upon which we focus and the four-point
correlation function of four-quark bilinear sources are related by the
standard Lehmann-Symanzik-Zimmermann (LSZ) reduction formula.  That
is, the scattering amplitude is given by the appropriately normalized
{\em amputated\/} four-point function.  For the case in which the
incident and final mesons are (pseudo)scalars, the four-point
correlation function in momentum space depends upon six kinematic
variables: $q_A^2$, $q_B^2$, $q_C^2$, and $q_D^2$, as well as $s$ and
$t$.  The scattering amplitude, in contrast, fixes $q_A^2$, $q_B^2$,
$q_C^2$, and $q_D^2$ to assume their on-shell values for the
particular mesons of interest, and is given by
%
%\begin{widetext}
%\begin{equation}
%\begin{split}
%  A(s,t) &=  \lim_{\begin{array}{c}q^2_A \rightarrow m_A^2\\q^2_B
%\rightarrow m_B^2 \\ q^2_C \rightarrow m_C^2 \\q^2_D \rightarrow m_D^2
%\end{array}}Z_A^{-1/2} Z_B^{-1/2}Z_C^{-1/2}Z_D^{-1/2}\, (q_A^2-m_A^2)
%(q_B^2-m_B^2)  (q_C^2-m_C^2) (q_D^2-m_D^2)  \, G_4^{ABCD}
%(q_A^2,q_B^2,q_C^2,q_D^2,s,t) \\  \\
%  Z_i &= | \lim_{q^2_i \rightarrow m_i^2} G_2^i (q^2_1)| \; \; \; \;
%  {\rm for} \; i=A,B,C,D \;
%\end{split}
%\label{Eq:disp}\end{equation}
%\end{widetext}
\begin{eqnarray}
A(s,t) & = & \prod_i \lim_{q_i^2 \rightarrow m_i^2} Z_i^{-\frac 1 2}
(q_i^2 - m_i^2) G_4^{ABCD} \left( \{ q_i^2 \}, s, t \right) ,
\nonumber \\
Z_i & = & \lim_{q_i^2 \rightarrow m_i^2} |G_2^i ( q_i^2 )| \ \ {\rm for}
\ i = A,B,C,D \, ,
\label{Eq:disp}
\end{eqnarray}
where $G_4^{ABCD}$ is the four-point correlation function in momentum
space for quark bilinear sources $J_{A,B,C,D}$, and $G_2^i$ is the
diagonal two-point correlation function for linear source $i$\@.  Note
that this structure guarantees the external mesons are always on shell
in the scattering amplitude.  However, the amplitude $G_4$ itself is
taken to be a general function of (complex) $s$ and $t$, which can
correspond to unphysical values that cannot be reached by the on-shell
mesons.  Analogous expressions hold for the case in which the external
mesons carry spin (Recall that the large-$N_c$ world supports stable
higher spin-mesons).

It is also useful to note that the scattering amplitude can be written
in the form of a fixed-$t$ dispersion relation:
%~\cite{****}:
%
\begin{equation}
  A(s,t) = a_1(t) + a_2(t) s + \frac{1}{\pi} \int d s' \frac{\rho(s',t)
s^2 }{s'^2(s-s'+ i \epsilon)} \label{Eq:dispersion}
\end{equation}
where we have assumed two subtractions.  We make the standard
assumption that this form continues to hold in the large $N_c$ limit.
The key to our analysis is that, if the Weinberg scenario is correct,
it follows from this standard dispersion analysis that the scattering
amplitude for meson-meson scattering amplitudes has $s$-channel
spectral strength concentrated at the position of the tetraquarks over
a region of width $N_c^{-1}$ and with integrated strength of
$O(N_c^{-1})$.  We focus on the $s$ channel since the quantum numbers
are given by the incident mesons, and the physical production of
tetraquarks in scattering occurs only in this channel due to kinematic
constraints.  If one can demonstrate that the spectral function
$\rho(s,t)$ for meson-meson scattering does not support $s$-channel
spectra of this character, then the Weinberg scenario is ruled out.

The remainder of this paper is organized as follows:
Sec.~\ref{sec:Diagrams} focuses on Feynman diagrams at the quark-gluon
level that contribute to the leading-order connected four-point
correlation function for quark bilinear sources.  A critical issue is
the connection between a spacetime representation of the diagrams and
a description in terms of color flow.  In fact, the Feynman diagrams
are actually identical and are merely drawn in different ways.
Section~\ref{sec:Cuts} identifies all possible types of cuts of these
leading-order connected diagrams and the intermediate states revealed
by them.  These cuts play a critical role in that they are associated
with the spectral function in Eq.~(\ref{Eq:dispersion}).  A careful
analysis of these cuts shows that no possible cut corresponds to a
tetraquark intermediate state in meson-meson scattering.  We summarize
and discuss our conclusion in Sec.~\ref{sec:Concl}; here, the issue of
cryptoexotics is addressed briefly.  The analysis depends upon a
number of standard assumptions used in large $N_c$, which are also
discussed briefly in the concluding section.
%In Appendix~\ref{sec:Standard}.

\section{Spacetime and Color-Flow Diagrams}
\label{sec:Diagrams}

\begin{figure*}
\begin{center}
\includegraphics[]{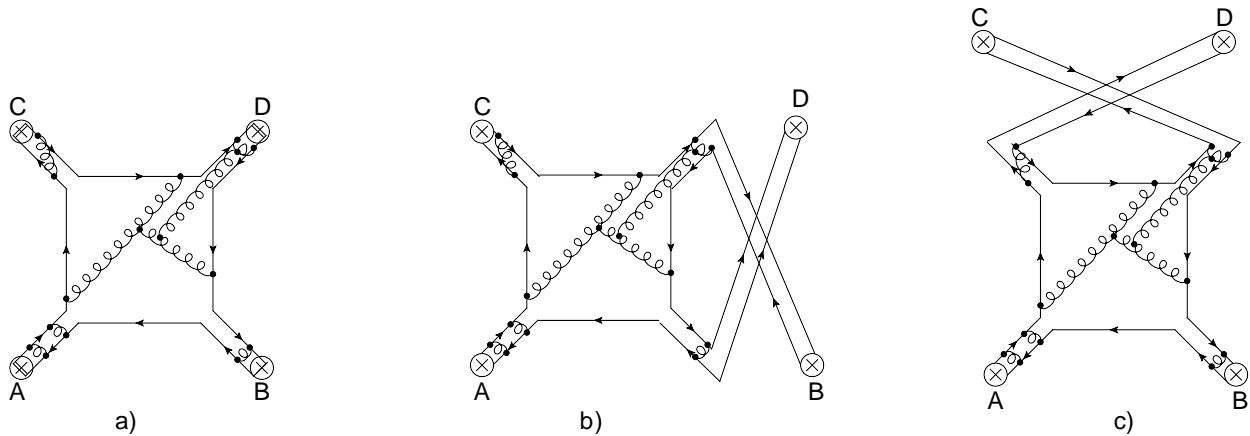}
\caption{Typical leading-order ($N_c^1$) Feynman diagrams for the
  connected four-point function.  In each case, the diagram follows a
  spacetime ordering, in that the bilinear sources $A,B$ represent
  initial states and the bilinear sources $C,D$ represent final
  states.}\label{Fig:typical_legs}
\end{center}
\end{figure*}

This section focuses on the four-point function of $\overline{q}q$
bilinear sources labeled $A, B, C, D$.  Following
Weinberg~\cite{Weinberg:2013cfa}, we consider only the leading-order
connected diagrams.  As a first step, we consider the diagrams
rendered in spacetime so that the initial meson sources are placed at
the bottom of the diagram and the final meson sources (actually sinks)
are placed at the top ({\it i.e.}, time flows upward).  In this
analysis, we focus on the $N_c$ counting and assume the standard
counting rules: We count only $N_c$ factors arising from color traces
(each closed color loop contributing $N_c^1$) and factors of $g_s \sim
N_c^{-1/2}$ from gluon vertices.

Typical leading-order connected diagrams are indicated in
Fig.~\ref{Fig:typical_legs}.  It is straightforward to show, using
't~Hooft double-line color-flow diagrams, that all of these diagrams are
$O(N_c^1)$ and hence of leading order.  This result might come
as something of a surprise since diagrams b) and c) do not appear to
be planar, and it is well known that all leading-order connected
diagrams containing quark lines are planar and bounded by the quark
loop.  However, the sense in which planarity holds has to do with
color flow and not with spacetime.  That is, a diagram is of leading
order in $N_c$ counting if, ignoring spacetime considerations, the
diagram {\em can\/} be drawn in a plane with the quark line bounding
the diagram.  Planarity is entirely determined by how the quarks and
gluons in the diagram are connected topologically and not on how
momentum flows from source to sink.  This distinction is critically
important since the question of $s$-channel spectral strength clearly
depends upon the spacetime flow of momentum in the diagram.

\begin{figure*}
\begin{center}
\includegraphics[]{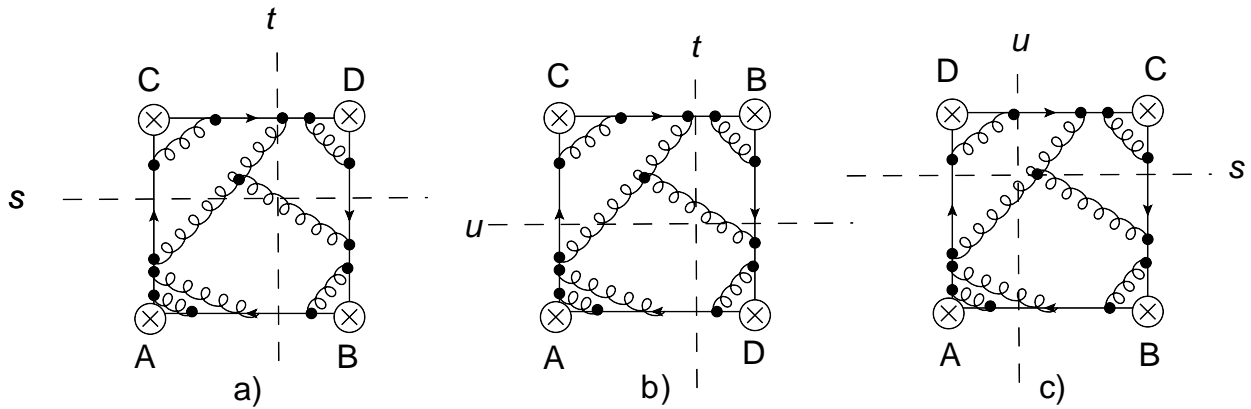}
\caption{The diagrams of Fig.~\ref{Fig:typical_legs}, recast as
  leading-order ($N_c^1$), manifestly planar diagrams in terms of
  color structure.  Cuts giving intermediate states in the indicated
  Mandelstam channels are denoted by dashed lines.}
\label{Fig:typical_cut}
\end{center}
\end{figure*}

The leading-order connected diagrams in Fig.~\ref{Fig:typical_legs}
can be easily seen to be planar and bounded by a quark loop if one
ignores the momentum flow and unfolds them, as depicted in
Fig.~\ref{Fig:typical_cut}.  In fact, the diagrams in
Fig.~\ref{Fig:typical_legs} are identical to the corresponding ones in
Fig.~\ref{Fig:typical_cut}, in which it is apparent that any
leading-order connected diagram associated with a four-point function
can be represented by a square with the four quark bilinear sources
(sinks) on the vertices.  One can then return to momentum-flow form by
considering how the sources inject momentum into the system.  Again,
the issue is topological: how the various momenta inserted into the
system are related.  One can divide such diagrams into topological
classes.  Superficially there appear to be 24 classes of leading-order
connected diagram since each of the four bilinears can attach to any
of the four vertices.  However, since the value of the diagram does
not change if the square is rotated through any multiple of $\frac \pi
2$ radians, the number of independent classes reduces to six.  One can
label the class of diagram by the order of the vertices, with $A$ and
$B$ representing the sources of the two incoming mesons and $C$ and
$D$ the outgoing ones, ordered counterclockwise starting at the bottom
left, so that diagram a) of Fig.~\ref{Fig:typical_cut} is labeled
$ABDC$.  Cyclic combinations of the square are identified because they
correspond to the same diagram: $ABDC$ is equivalent to $BDCA$.
Finally, because
% the interaction is time-reversal invariant,
of crossing symmetry, if one reverses the order of the sources and
flips all internal gluons along the diagonal, the value of the diagram
% does not change.
changes by at most a phase.  Thus one can reverse the ordering so
that, {\it e.g.}, $ABDC$ is in the same class as $ACDB$, leaving three
classes.  Figures~\ref{Fig:typical_legs} and \ref{Fig:typical_cut}
give one representative of each class.

It is worth noting that not all of these classes of leading-order
connected diagrams contribute to processes in the quantum-number
exotic channels of interest here.  Consider for concreteness the case
of processes that
%isospin 2 processes---which means that the process must
carry $I=2$ in the $s$ channel.  It is easy to show that classes
$ABDC$ and $ABCD$ [corresponding to diagrams a) and c) in
  Fig.~\ref{Fig:typical_cut}] do not contribute to $I=2$ processes in
the $s$ channel, while class $ADBC$ [corresponding to diagram b) in
  Fig.~\ref{Fig:typical_cut}] does.  To see why, note that the $s$
channel carries the quantum numbers injected at sources $A$ and $B$\@.
Since the process carries $I=2$ in the $s$ channel, the sources $A$
and $B$ together must provide $I=2$.  Moreover, since the gluons are
isosinglets, isospin can be injected into the diagram only at sources
$A$, $B$, $C$, and $D$\@.  In diagrams of classes $ABDC$ and $ABCD$,
vertices $A$ and $B$ are adjacent.
%If one follows the quark line
%connecting the adjacent $A$ and $B$ vertices, the $I\le\frac 1 2$ quarks
%interacts with a combined $I=2$ operator associated with vertices.
%This must vanish by the Wigner-Eckart theorem; it is proportional to
%a diagonal matrix element of an $I=2$ operator between $I=\frac 1 2$
%states.
The quark line connecting the adjacent $A$ and $B$ vertices, treated
as a $\overline{q}q$ pair, can be considered an isosinglet source
(since deep in the diagram, it turns into pure glue); the two remaining
quarks in the $s$-channel can carry no more than $I=1$ and therefore
give a vanishing matrix element with an $I=2$ operator.

\section{Cuts and intermediate States}
\label{sec:Cuts}

This section describes the role in the analysis of ``cuts'', for which
two related meanings are relevant.  The first meaning is related to
the analytic structure of the scattering amplitude [$A(s,t)$ in
Eq.~(\ref{Eq:disp})], or more generally, $n$-point correlation
functions $G_n$.  Consider for example Eq.~(\ref{Eq:disp}), which
relates the real and imaginary parts of $A(s,t)$.  As an analytic
function of $s$, $A(s,t)$ has a cut over the full region in which the
spectral function $\rho(s,t)$ is a nonzero and continuous function.
Note that the cut is associated with accessible physical states: It
corresponds to all values of $s$ for which the system can go on shell.

At the level of perturbation theory for the underlying theory of
quarks and gluons, ``cut'' can also be used to describe the process of
dividing a diagram into two distinct regions.  Typically, this
separation can be done in more than one way.  The region along the
boundary between the regions can be referred to as the ``cut''.  The
combination of quarks and gluons along the cut corresponds to a
particular quark-gluon intermediate state contributing in perturbation
theory.

These two meanings are strongly related: When the hadronic
intermediate states revealed by cutting a diagram go on shell, they
contribute to the cut in the sense of the analytic function (to the
extent that such states are calculable in perturbation theory).

One class of cuts is easily depicted by using the color-flow diagrams
of Fig.~\ref{Fig:typical_cut}: A diagram in each of the three classes
can be cut in two natural ways.  In the class $ABDC$ of a), one can
generate an $s$-channel cut (cutting through both quark lines
separating the $A$ and $B$ vertices from the $C$ and $D$ vertices) or
a $t$-channel cut (separating vertices $A$, $C$ from $B$, $D$).  Of
course, more than one particular cut is available, depending upon
which specific gluon lines are cut, but these two basic cut structures
naturally arise.  Similarly, Fig.~\ref{Fig:typical_cut} shows $u$- and
$t$-channel cuts for class $ADBC$ and $s$- and $u$-channel cuts for
class $ABCD$.

The types of cuts shown in Fig.~\ref{Fig:typical_cut} clearly do not
contain intermediate quark-gluon states that contribute to tetraquark
states.  All of these cuts contain precisely one $q\overline{q}$ pair
in the intermediate states; thus, they contribute to mesons, not
tetraquarks.  Recall that we seek only $s$-channel singularities
connected with the physical intermediate states created in the
scattering process.  It is significant that Fig.~\ref{Fig:typical_cut}
has $s$-channel cuts only in classes $ABDC$ and $ABCD$ but not $ADBC$,
but this result is not surprising: As noted in
Sec.~\ref{sec:Diagrams}, $s$-channel quantum-number exotic channels
correspond only to class $ADBC$ among the leading-order connected
diagrams.  If class $ADBC$ had meson-type $s$-channel contributions,
one would have a contradiction: The exotic channels associated with
this class by definition lack ordinary meson contributions.

\begin{figure*}
\begin{center}
\includegraphics[]{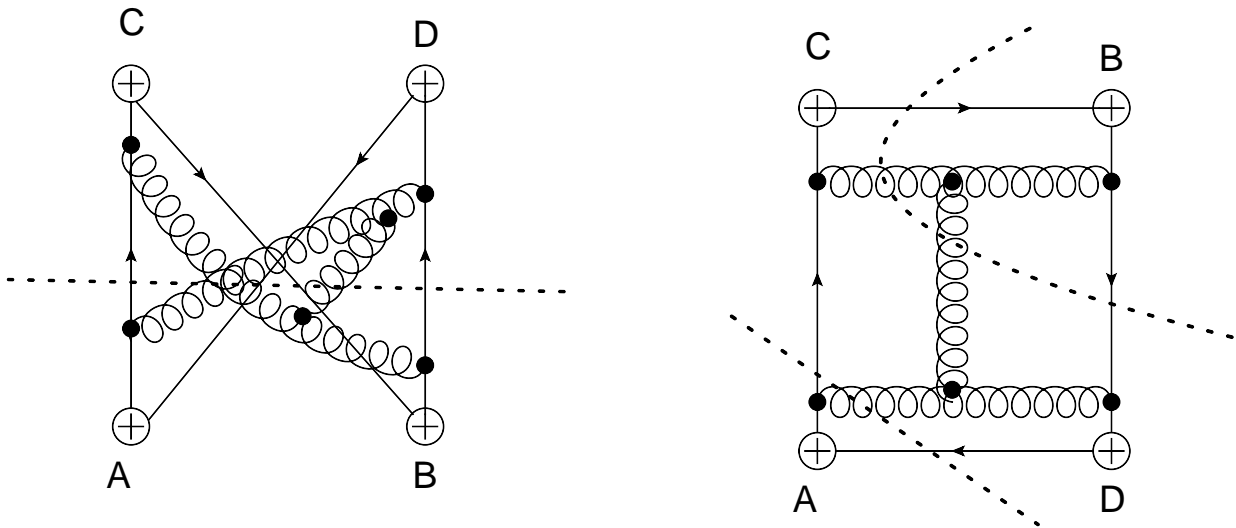}
\caption{An $s$-channel cut in leading-order ($N_c^1$) drawn both as
  a spacetime diagram and as a manifestly planar diagram emphasizing
  the color structure.}
\label{Fig:topprop}
\end{center}
\end{figure*}

However, as it turns out, the kinds of cuts shown in
Fig.~\ref{Fig:typical_cut} that seem natural when the diagrams are
drawn in planar form are not the only ones appearing in the four-point
functions at leading order in $N_c$.  The key point is that the
diagrams of Fig.~\ref{Fig:typical_cut} are drawn in a form to
emphasize color-flow connectivity and not the flow in spacetime.
However, the spacetime description is the most relevant one for
understanding the cuts.  To make this point clear, consider
Fig.~\ref{Fig:topprop}.  The diagram is of class $ADBC$ and hence can
arise in an exotic $s$-channel scattering process.  This class is also
analogous to diagrams considered by Weinberg in
Ref.~\cite{Weinberg:2013cfa}.  The critical point is that the diagram
viewed in its spacetime form clearly has an $s$-channel cut passing
through two quarks and two antiquarks.  That is, it has the correct
quantum-number content to be a tetraquark.  In fact, if one considers
quantum-number exotic channels and restricts attention to
leading-order connected diagrams, all $s$-channel cuts clearly have
this character: The intermediate states revealed by the cut always
contain two quarks and two antiquarks.

The critical question is whether these cuts reveal resonant tetraquark
states or merely two noninteracting (or weakly interacting) mesons.
It is possible to definitively answer this question if one accepts the
standard assumptions used in large $N_c$ analysis.  To see how, note
that if one unfolds the diagram of Fig.~\ref{Fig:topprop} given in
spacetime form (the first inset) to write it in manifestly planar
color-flow form (the second inset), the line denoting the cut breaks
into two distinct parts.  In the case indicated here, one part cuts
off vertex $A$ and the other cuts off vertex $B$; that is, it
literally cuts the corners off the diagram (In the literal
transcription of this cut, the cut termini at the top and bottom of
the color-flow diagram of Fig.~\ref{Fig:topprop} are identified).  It
is guaranteed that {\em any\/} $s$-channel cut for all diagrams in
class $ADBC$ (and hence for all leading-order connected diagrams in
quantum-number exotic channels) cut corners in the diagram written in
planar form, either the corners $A$ and $B$ as the preceding example,
or the corners $C$ and $D$\@.  The reason is simple topology: In class
$ADBC$, corners $A$ and $B$ are opposite from each other, as are
corners $C$ and $D$\@.  However, an $s$-channel cut by definition cuts
the region associated with the incoming particles (which contains
vertices $A$, $B$) from the region associated with the outgoing
particles (which contains vertices $C$, $D$).  In a spacetime
description, a single line across the diagram always accomplishes the
cut.  However, since $A$ and $B$ lie diagonally across from each other
in planar form (as do $C$ and $D$), no single line across the diagram
can represent the cut.  The only resolution is to make such a cut
remove two diagonally separated corners and associate the disconnected
regions of the two corners to be part of the same ``side'' of the cut.

The physical significance of the fact that the $s$-channel cuts always
isolate opposite corners for leading-order connected diagrams in
exotic channels (and for class $ADBC$ generally) is clear: The
individual cuts across the two corners carry precisely the
four-momentum and color structure (singlet) injected at each of the
two sources.  The four-momentum entering at sources $A$ and $B$ is
conserved at every vertex and must flow out of the corner, but the
only way out of the corner is across the cut.  This result is crucial
because the object of interest is the scattering amplitude
$A(s,t)$---{\it i.e.}, the appropriately normalized {\em amputated\/}
four-point correlation function---rather than the four-point
correlation function $G_4^{ABCD}$ itself; as noted in the
Introduction, the scattering amplitude {\em must\/} contain a
tetraquark if the mesons couple to tetraquarks.  Moreover, if one
accepts the standard assumptions underlying the large $N_c$ analysis
of hadrons and Weinberg's scenario that tetraquarks couple to two
meson with a strength of $\sim N_c^{-1/2}$, then it must be true that
the scattering amplitude computed using the leading-order connected
diagrams contains analytic singularities in the $s$ channel associated
with the tetraquark intermediate state.  However, the LSZ reduction
rule removes all spectral strength from the incoming and outgoing
mesons to produce the scattering amplitude.  In $A(s,t)$, the only
contributions to the spectral strength $\rho(s,t)$ come from
intermediate states that are distinct from the incident and outgoing
ones; the factors of $(q_i^2-m_i^2)$ in Eq.~(\ref{Eq:disp}) ensure
this.  In particular, when one cuts the corner of a diagram, the
contribution associated with the cut corner to the scattering gives no
spectral strength and thus no cut (in the sense of analytic
functions); it has been removed from the full four-point function via
LSZ reduction.  Physically, the spectral strength in the four-point
function associated with cut corners corresponds to the external meson
states only and not to any intermediate state.

The preceding argument implies that an exotic diagram such as
Fig.~\ref{Fig:topprop} produces no spectral strength at leading $N_c$
order in the $s$ channel, and hence no leading-order contribution to
the scattering amplitude.  In terms of Eq.~({\ref{Eq:dispersion}}),
$\rho(s,t)$ vanishes at $O(N_c^{-1})$.  However, in the scenario in
which tetraquarks exist and couple to two mesons with a strength $\sim
N_c^{-1/2}$, tetraquarks must be present in the scattering amplitude
if the standard assumptions for large $N_c$ hadrons hold.  Thus we are
forced to conclude that, for the case of exotic channels, either the
scenario in which quantum-number exotic tetraquarks exist and couple
with a strength $\sim N_c^{-1/2}$ is incorrect, or the standard large
$N_c$ assumptions are inadequate.

We should note that the vanishing of the leading-order spectral
strength in Eq.~({\ref{Eq:dispersion}}) for exotic channels does {\em
  not\/} imply the absence of scattering at this order.  Rather, it
means that leading-order scattering comes through $a_1(t)$ and
$a_2(t)$, which can arise through $t$-channel exchange.

%%%%%%%%%%%%%%%%%%%%%%%%%%%%%%%%%%%%%%%%%%%%%%%%%%

\section{Conclusions} \label{sec:Concl}

The argument we have presented implies that either the scenario of
Ref.~\cite{Weinberg:2013cfa}, in which quantum-number exotic
tetraquarks exist and couple with a strength $\sim N_c^{-1/2}$, is
incorrect, or the standard assumptions about hadrons in large $N_c$
are inadequate.

Let us first consider the possibility that some of the standard
assumptions used in large $N_c$ hadronic analysis might be incorrect.
One should note that a variety of such assumptions occur---some
explicitly made in the large $N_c$ literature and others that are
implicit.  We believe that these assumptions are highly plausible and
we have no reason to doubt any of them.

As a question of logic, however, it remains possible that one or more
of these assumptions is not valid and that as a result, quantum-number
exotic tetraquarks are not excluded from arising through leading-order
connected diagrams.  Let us consider an example: It is always
implicitly assumed that the standard dispersion relations and spectral
representations of field theory hold in the large $N_c$ limit.  Now of
course, they are supposed to hold for any legitimate field theory, any
hence should hold for QCD at any $N_c$.  However, this assertion does
not necessarily imply that these relations are valid for the large
$N_c$ limit of the theory.  Such issues could arise, for example, due
to some types of nonanalytic behavior in $1/N_c$ for correlation
functions as $N_c \rightarrow \infty$.  As a concrete model in which
such behavior could allow quantum-number exotic tetraquarks, consider
the example in Ref.~\cite{Lebed:2013aka}.  There, it is assumed that
the four-point function in position space has a coupling to the
tetraquark scaling as $\exp[-N_c^{1/3} (x_1-x_2)^2]$, where $x_1$ and
$x_2$ are the positions of the two quark bilinears.  Such nonanalytic
behavior has the consequence that widely separated quark bilinear
sources---the type associated with external mesons---do not couple to
tetraquarks, while overlapping sources do.  Thus, even though
tetraquarks by construction exist in the theory and couple to mesons,
at large $N_c$ they have no spectral strength in the scattering
amplitude, in contradiction to what one expects from a spectral
representation.

The example above is hardly unique.  Many assumptions used in large
$N_c$ analysis are believed to be true but have never been rigorously
proven and, if false, would invalidate the argument of the preceding
section.  For example, a basic tenet of the analysis to assume the
leading $1/N_c$ scaling of a correlation function is the same as the
that of the leading class of contributing Feynman diagrams.  This
assertion is highly plausible and is the basis for almost all analyses
of hadrons at large $N_c$.  It is nevertheless not guaranteed to be
true in a mathematical sense, since the perturbative expansion
embodied in the Feynman diagrams is not convergent.  If it is false,
one can easily evade the argument given above: Only the {\em
  leading}-order diagrams lack spectral strength in the $s$ channel
for the scattering amplitude in exotic channels, and the tetraquark
could be built up from formally subleading perturbative orders.
Overall, we view such possibilities (while being logically not
excluded) as being highly unlikely.  The standard large $N_c$ analysis
has produced many useful insights into hadronic physics, and it seems
unlikely that exceptions should first become apparent in the
tetraquark sector.
% To focus on the possibility that some of the assumptions underlying
% this analysis somehow happen to be wrong for tetraquark channels
% seems like nitpicking.

If one accepts the standard large $N_c$ assumptions, then the analysis
here rules out the possibility that quantum-number exotic channels
follow the scenario of Ref.~\cite{Weinberg:2013cfa}, in which
tetraquarks exist and arise due to the leading-order connected
diagrams, and accordingly couple to two mesons with a strength of
$O(N_c^{-1/2})$.  Then two possibilities remain: Either narrow
tetraquarks do not exist, or they arise from subleading diagrams and
couple to mesons at $O(N_c^{-1})$ or less (making them even narrower).

Consideration of the second possibility actually follows the basic
insight Ref.~\cite{Weinberg:2013cfa}, where Weinberg observed
that just because the leading [$O(N_c^2)$] contribution to the
two-point correlation function of a source with tetraquark quantum
numbers is a pure two-meson state with no resonant tetraquark
contribution does not mean that no tetraquark contribution arises in
the first subleading contribution, which happens to be the leading
contribution for the connected correlator.  Here we have shown (modulo
the assumptions discussed above) that, in exotic channels the leading
connected diagrams also contain no tetraquark contributions.  However,
a simple extension of Weinberg's logic leads one to the observation
that subleading contributions could contain a tetraquark.

How plausible is this?  We believe it unlikely to be correct.  To see
why, consider the analogous situation of mesons at large $N_c$, in
which mesons arise from the leading-order [$O(N_c^1)$] correlation
functions of quark bilinears.  One deduces that mesons, for reasons of
self-consistency, have widths that scale as $1/N_c$.  Using the
standard analysis, can one falsify the existence of an additional
class of special mesons with the same quantum numbers as the ordinary
ones but that arise from subleading diagrams and have anomalously
small [$O(1/N_c^2)$] widths?  At a strict formal level, the answer is
``no''.  However, the idea seems quite far-fetched.  Indeed, Occam's
razor argues against it: We know of no reason why such a peculiar
behavior ought to emerge, and in the absence of a compelling reason,
it appears to be unlikely.  In much the same way, it seems quite
far-fetched that ultra-narrow tetraquarks should arise from subleading
diagrams.  As discussed in Ref.~\cite{Knecht:2013yqa}, one finds
channels with special flavor content that would require tetraquarks
(were they to exist) to have particularly narrow [$O(N_c^{-2})$]
widths.  However, the case considered here is whether ``ordinary''
tetraquarks with, {\it e.g.}, $I=2$ are anomalously narrow.

Thus, we conclude that quantum-number exotic narrow tetraquarks are
unlikely to exist as resonant states at large $N_c$ in QCD(F).  This
behavior is qualitatively very different from that of QCD(AS), in
which such states are known to exist~\cite{Cohen:2014via}.  Such a
difference is both awkward and interesting, since one hopes to use
large $N_c$ behavior as a guide for our $N_c=3$ world.  Since the two
distinct large $N_c$ limits are qualitatively different for
tetraquarks, the presence or absence of observably narrow tetraquarks
at $N_c=3$ provides an exciting opportunity to distinguish which limit
is superior in this case.

Finally, the analysis here is limited to quantum-number exotic
channels.  What would change for nonexotic channels?  As noted above,
the situation becomes more ambiguous, because it is difficult to
separate cryptoexotic components from those of conventional mesons.
One thing is clear: Our analysis shows that the diagrams in class
$ADBC$ cannot contribute to the spectral strength in the scattering
amplitude for tetraquark channels.  On the other hand, as seen in
Fig.~\ref{Fig:typical_cut}, classes $ABDC$ and $ABCD$ do have
$s$-channel spectral strength, but these are of a $\overline{q}q$
nature and can be ascribed to mesons rather than tetraquarks.  Thus it
appears likely that any sensible definition of ``tetraquark'' in
cryptoexotic channels yields is no specifically tetraquark spectral
strength.

%%%%%%%
%Should a narrow tetraquark appear in QCD(F) at large $N_c$, it would
%appear to require some $N_c$ factors to emerge from nonconventional
%sources, as in Ref.~\cite{Lebed:2013aka}.
%%%%%%%

%%%%%%%
%However, to interpret $J$, which is factorizable as $BB$ and therefore
%can create two mesons with $f_M = O(N_c^{1/2})$ each, as a
%single-tetraquark source with $f_T = O(N_c^{1/2})$ requires a subtle
%limiting process in $N_c$ when the two separated $B$ sources coalesce
%into a local $J$ operator~\cite{Lebed:2013aka}.
%%%%%%%

Since the large $N_c$ limit of QCD(F) has produced so many useful
phenomenological insights, the exciting possibility of narrow
tetraquarks gives hope for the eventual observation of such a state.
However, we have found that no mechanism relying upon the conventional
counting of $N_c$ factors, even ones with unusual diagrammatic and cut
structures, give rise to parametrically narrow tetraquarks.  Even so,
the alternative and phenomenologically viable QCD(AS) large $N_c$
limit remains a possibility for producing naturally narrow
tetraquarks.  Lastly, other suppressions having nothing to do with
$N_c$ counting ({\it e.g.}, by heavy quark masses, as may be the case
for the $X(3872)$~\cite{Beringer:1900zz}) might still make an
observably narrow tetraquark state viable.

\begin{acknowledgments}
This work was supported by the U.S.\ Department of
Energy under Grant DE-FG02-93ER-40762 (T.D.C.) and by the National
Science Foundation under Grant PHY-1068286 (R.F.L.).
\end{acknowledgments}

\end{document}